# Autoamplification and competition drive symmetry breaking: Initiation of centriole duplication by the PLK4-STIL network

Marcin Leda[1], Andrew J. Holland[2], Andrew B. Goryachev[1,*]

[1]Centre for Synthetic and Systems Biology, School of Biological Sciences, University of Edinburgh, Edinburgh, EH9 3BF UK

[2]Department of Molecular Biology and Genetics, Johns Hopkins University School of Medicine, Baltimore, Maryland 21205, USA

*Correspondence: Andrew.Goryachev@ed.ac.uk

## Summary

Symmetry breaking, a central principle of physics, has been hailed as the driver of self-organization in biological systems in general and biogenesis of cellular organelles in particular, but the molecular mechanisms of symmetry breaking only begin to become understood. Centrioles, the structural cores of centrosomes and cilia, must duplicate every cell cycle to ensure their faithful inheritance through cellular divisions. Work in model organisms identified conserved proteins required for centriole duplication and found that altering their abundance affects centriole number. However, the biophysical principles that ensure that, under physiological conditions, only a single procentriole is produced on each mother centriole remain enigmatic. Here we propose a mechanistic biophysical model for the initiation of procentriole formation in mammalian cells. We posit that interactions between the master regulatory kinase PLK4 and its activator-substrate STIL form the basis of the procentriole initiation network. The model faithfully recapitulates the experimentally observed transition from PLK4 uniformly distributed around the mother centriole, the "ring", to a unique PLK4 focus, the "spot", that triggers the assembly of a new procentriole. This symmetry breaking requires a dual positive feedback based on autocatalytic activation of PLK4 and enhanced centriolar anchoring of PLK4-STIL complexes by phosphorylated STIL. We find that, contrary to previous proposals, *in situ* degradation of active PLK4 is insufficient to break symmetry. Instead, the model predicts that competition between transient PLK4 activity maxima for PLK4-STIL complexes explains both the instability of the PLK4 ring and formation of the unique PLK4 spot. In the model, strong competition at physiologically normal parameters robustly produces a single procentriole, while increasing overexpression of PLK4 and STIL weakens the competition and causes progressive addition of procentrioles in agreement with experimental observations.

## Introduction

Symmetry breaking, an overarching principle of modern physics, explains the emergence of new order in initially disordered systems and has long been suggested to drive self-organization of biological systems, albeit very few specific examples have been elucidated (Goryachev and Leda, 2017; Kirschner et al., 2000). Thus, biogenesis of membraneless cellular organelles, such as centrosomes and nucleoli, has recently been proposed to represent nonequilibrium phase separation (Hyman et al., 2014), a particular realization of symmetry breaking on the intracellular scale. While the abstract principle of symmetry breaking is generally well accepted in biology, by itself it does not help biologists to understand specific experimental observations. Biophysical models that translate apparently complex molecular mechanisms into transparent physical principles are necessary to bring our understanding of cellular morphogenesis to the new qualitative level. Here we apply theoretical modeling to propose that duplication of centrioles is a manifestation of symmetry breaking driven by autoamplification and competition.

Centrioles play vital cellular roles in regulating cell division as the cores of centrosomes and in ciliogenesis as the precursors of cilia basal bodies (Lattao et al., 2017; Loncarek and Bettencourt-Dias, 2017; Nigg and Stearns, 2011). These submicron-sized membraneless organelles have cylindrical shape and intricate molecular architecture with an unusual ninefold rotational symmetry (Gonczy, 2012; Jana et al., 2014). The enigmatic biogenesis of centrioles has been a matter of much interest since their discovery in the late 19$^{th}$ century (Banterle and Gonczy, 2017; Marshall et al., 2001). Somatic eukaryotic cells inherit two centrioles from their mothers and each of these centrioles must duplicate precisely once per cell cycle to ensure that the cell's daughters receive exactly two centrioles again. Both the failure to duplicate and the production of supernumerary centrioles can lead to genomic instability and cellular death. Therefore, errors in the numeric control of centriole biogenesis are associated with human diseases, such as microcephaly and cancer (Gonczy, 2015; Levine et al., 2017; Marthiens et al., 2013; Nigg and Holland, 2018). Much has been learnt recently about the molecular mechanisms of temporal control that ensure that the replication process is initiated only once per cell cycle (Loncarek et al., 2010; Novak et al., 2016; Shukla et al., 2015; Tsou and Stearns, 2006; Wang et al., 2011). However, how precisely one procentriole is formed per mother centriole within one round of replication, i.e. the numeric control, is still far from being understood.

Considerations of symmetry are important for understanding mechanisms of biological replication. Many axisymmetric unicellular organisms, such as bacteria and fission yeast, replicate by first growing along the axis of symmetry and then pinching in two. As centrioles are axially symmetric, it would seem logical that their duplication could be most easily achieved by such a mechanism, templated extension followed by division. Contrary to these naïve expectations, early microscopy studies revealed that the procentriole is formed at the base of mother centriole so that their axes are perpendicular to each other (Schreiner and Schreiner, 1905). This unexpected spatial arrangement suggests a mechanism inconsistent with a simple template-extension scenario. Furthermore, under some circumstances, centrioles can form *de novo*, away from any pre-existing centrioles (Khodjakov et al., 2002; Marshall et al., 2001). The relative roles of self-organization versus templated growth have been extensively discussed in the literature (Karsenti, 2008; Rodrigues-Martins et al., 2007), however, the detailed understanding of centriole replication mechanisms began to emerge only recently, with the elucidation of involved molecular players and their mutual interactions.

Studies in worm *C. elegans*, fly *D. melanogaster*, and mammals revealed a core set of functionally conserved interacting proteins that are required for centriole replication. Serine-threonine protein kinase PLK4, a member of the polo-like kinase family (Zitouni et al., 2014), has emerged as the master regulator of procentriole biogenesis. Local activation of PLK4 at the base of mother centriole has been shown to be absolutely essential for the procentriole initiation, growth and number control (Aydogan et al., 2018; Bettencourt-Dias et al., 2005; Habedanck et al., 2005; Ohta et al., 2018; Pelletier et al., 2006). Importantly, overexpression of PLK4 and other proteins from the core replication set produces simultaneous formation of supernumerary procentrioles arranged around the base of mother centriole in a characteristic florette pattern (Kleylein-Sohn et al., 2007). Recent super-resolution microscopy analyses demonstrated that, even without overexpression, PLK4 first encircles the base of mother centrioles in a symmetric ring-shaped pattern but then undergoes a mysterious transformation into a single spot-like focus that eventually develops into the procentriole (Dzhindzhev et al., 2017; Kim et al., 2013; Ohta et al., 2014; Ohta et al., 2018). Thus, in contrast to the cell cycle-based temporal regulation of replication, numeric control of replication must involve spatial mechanisms. As PLK4 autophosphorylation leads to ubiquitylation and degradation of the kinase (Cunha-Ferreira et al., 2009; Holland et al., 2010; Rogers et al., 2009), it has been suggested that rapid degradation is responsible for the transformation of the ring into the spot (Ohta et al., 2014; Ohta et al., 2018). However, it remains unclear why and how degradation would favor a particular site to become the procentriole.

We performed integrative analysis of the existing cell biological, structural and biochemical data to propose a biophysical model of the early stages of procentriole formation. This model predicts that a single focus of PLK4 activity results from the breaking of symmetry of the spatially uniform ring state. We show that degradation, while important for maintaining low copy numbers of key proteins, by itself is insufficient to break the ring symmetry. Instead, the ability of PLK4 complexes to change their position on the surface of mother centriole by unbinding from one locus and re-binding at another is shown to be required for the symmetry breaking. Effectively, spatial loci on the ring compete for the PLK4 complexes and a single focus emerges as the winner of this competition. While the single focus is remarkably stable within a range of protein concentrations controlled by degradation, the model shows that further overexpression of the core proteins results in formation of supernumerary procentrioles in a characteristic dose-dependent pattern in full agreement with experimental results.

## Results and Discussion

### A model of centriole biogenesis

It has been established that three evolutionary conserved proteins are absolutely necessary for the initial stages of procentriole formation: the kinase PLK4/Zyg-1/Sak, scaffold protein STIL/Sas-5/Ana2, and the building block of the ninefold-symmetric cartwheel, SAS6 (Arquint and Nigg, 2016). Since these three proteins almost simultaneously appear at the site of the nascent procentriole and precede all others, we use modeling to explore the hypothesis that these key proteins are, in fact, sufficient for the induction of procentriole formation. Mammalian centrioles duplicate in early S phase of the cell cycle when PLK4, STIL and SAS6 are re-expressed after they had been degraded at the end of previous mitosis (Arquint et al., 2012; Sillibourne et al., 2010; Tang et al., 2011). We focus on the dynamics of PLK4

(P) and STIL (S) and do not consider SAS6 explicitly to reduce complexity of the model. The diagram of all model species and reactions is shown in Figure 1. All binding reactions are reversible, and the cytoplasmic species are denoted by the subscript *c*.

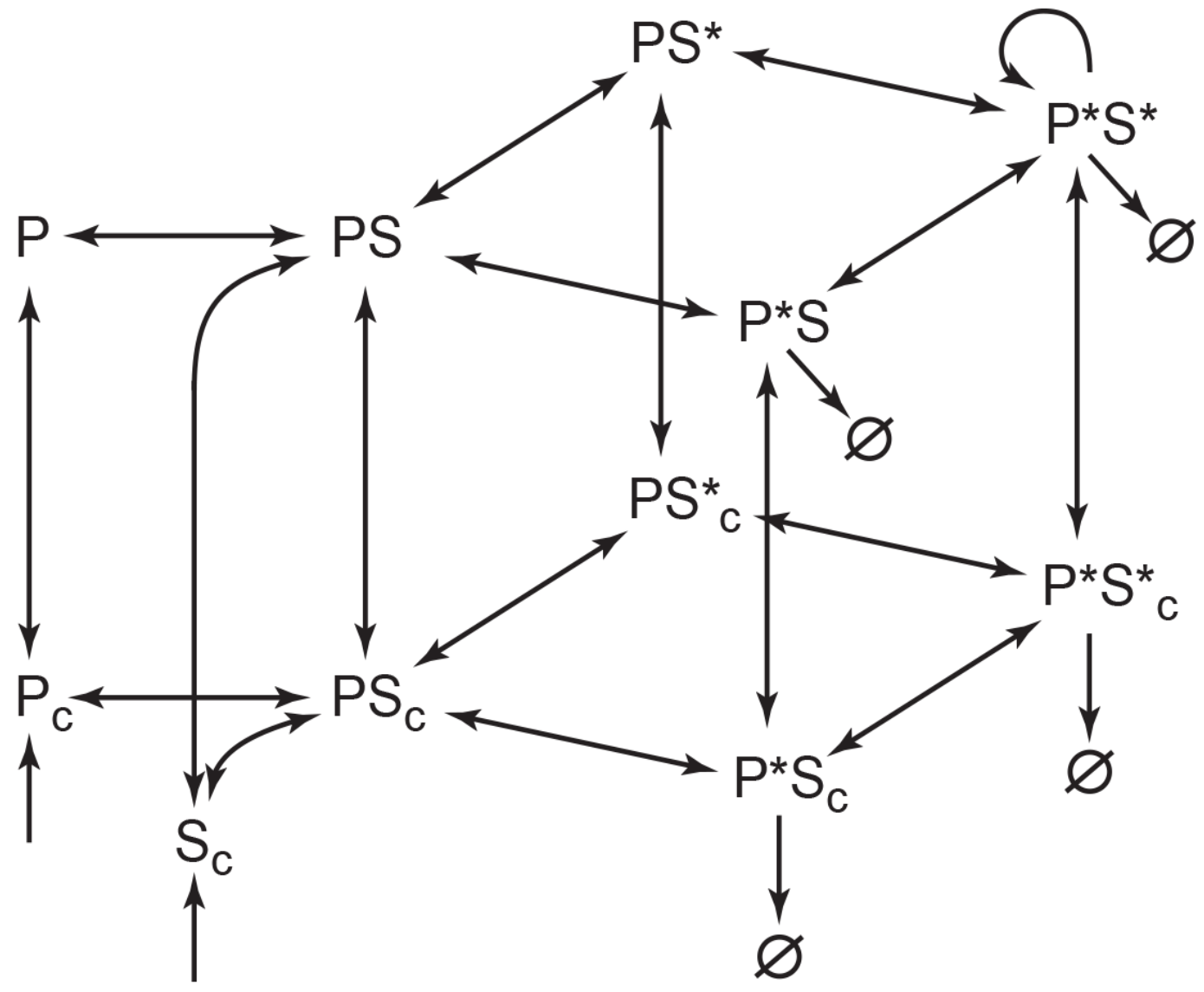

**Figure 1. A model of the reaction network proposed to initiate procentriole formation.** Kinase PLK4 (P), scaffold STIL (S) and their complexes are shown as the centriole-bound (top layer) and cytoplasmic species (bottom layer, denoted by subscript *c*). Asterisks represent phosphorylated species. Arrows show directionality of reactions, autocatalytic generation of P*S* is shown by the circular arrow, Ø denotes degradation of protein species. Weak degradation of unphosphorylated $P_c$ and $S_c$ is not shown. See also Figure S1.

Mammalian PLK4 begins to accumulate as an inactive kinase in G1 and is recruited to the surface of the pre-existing mother centrioles (henceforth, simply centrioles for brevity) via the well-characterized interactions with the scaffolds CEP192/Spd2 and CEP152/Asl (Cizmecioglu et al., 2010; Hatch et al., 2010; Kim et al., 2013; Sonnen et al., 2013). Rapid recovery of PLK4 fluorescence after photobleaching (Cizmecioglu et al., 2010) suggests that binding to centrioles is labile, yet sufficiently strong to provide the initial accumulation of PLK4 as a ring surrounding the proximal end of the centriole. As the newly synthesized STIL starts to accumulate in the cytoplasm in early S phase, it binds to the centriole-associated PLK4 (Arquint et al., 2015; Moyer et al., 2015; Ohta et al., 2014). This reaction initiates activation of PLK4 by relieving its intramolecular inhibition (Arquint et al., 2015; Klebba et al., 2015; Moyer et al., 2015; Ohta et al., 2018). We assume that allosteric activation and autophosphorylation of the activation loop of PLK4 (Klebba et al., 2015; Lopes et al., 2015) occur very rapidly upon STIL binding and, therefore, the complex of PLK4 and STIL (PS) contains active PLK4. PLK4 then sequentially autophosphorylates on multiple sites that include the degron motif, whose phosphorylation results in ubiquitination and subsequent rapid degradation of PLK4 (Cunha-Ferreira et al., 2009; Guderian et al., 2010; Holland et al., 2010; Peel et al., 2012; Rogers et al., 2009; Sillibourne et al., 2010). PLK4 also multiply phosphorylates the STIL molecule that it is bound to (Dzhindzhev et al., 2017; McLamarrah et al., 2018). This phosphorylation is important for the retention of STIL at the centriole (Ohta et al., 2018). Phosphorylation of the C-terminal STAN motif of STIL is required for the interaction between STIL and SAS6 (Dzhindzhev et al., 2014; Kratz et al., 2015; Moyer et al., 2015; Ohta et al., 2014). This binding is necessary for either the in-situ assembly or anchoring of the elsewhere preassembled SAS6 cartwheel (Fong et al., 2014). Phosphorylation of STIL by PLK4 is not restricted solely to the STAN motif and additional phosphosites, e.g., at the N-terminus (Dzhindzhev et al., 2017; Dzhindzhev et al., 2014; McLamarrah et al., 2018), may be important for the interaction of STIL with other centriolar proteins. Multiple phosphorylated species of the PLK4-STIL complex are represented in our model by the following four variables: PS (PLK4 phosphorylated only on the activation loop), P*S (fully phosphorylated PLK4), PS* (phosphorylated STIL), and P*S* (fully phosphorylated PLK4 and STIL). Transitions between these species are made reversible by the implicit action of several protein phosphatases (Brownlee et al., 2011; Kitagawa et al., 2011; Peel et al., 2017; Song et al., 2011; St-Denis et al., 2016; Wu et al., 2008).

Since PLK4 is a dimer, the two kinase domains are thought to phosphorylate the T-loop and PLK4 phosphodegron in trans, but still within the same PLK4-STIL complex (Guderian et al., 2010). Thus, these reactions can take place even at very low PLK4-STIL concentrations, such as those reported for the cytoplasm (Bauer et al., 2016). It has been proposed, however, that *Drosophila* PLK4 can promote its own activation in a concentration-dependent manner (Lopes et al., 2015). This result implies that PLK4 can also phosphorylate targets on other PLK4-STIL complexes that are in close physical proximity. Henceforth we refer to this type of PLK4 activity as crossphosphorylation. For the PLK4-STIL complexes to be able to crossphosphorylate on the surface of centriole, it would be necessary that they have i) high spatial density and ii) long residence time. Both requirements can be satisfied by the same molecular mechanism. Indeed, multiple lines of evidence indicate that phosphorylation of STIL by PLK4 increases centriolar retention of STIL-PLK4 complexes (Lambrus et al., 2015; Moyer et al., 2015; Ohta et al., 2018; Vulprecht et al., 2012; Zitouni et al., 2016). This PLK4 activity-dependent anchoring effect is likely to be mediated largely by the interaction of STIL with SAS6 complexes, but also could be enhanced by the interactions of phosphorylated STIL with other centriolar proteins and microtubules (Bianchi et al., 2018; Ohta et al., 2018). Therefore, we postulate the existence of a positive feedback loop in which PLK4 activity auto-amplifies itself by strengthening its centriolar anchoring and, therefore, increasing its spatial density. The increase in spatial density, in turn, results in stronger crossphosphorylation. In our model, this positive feedback is formulated as two assumptions. Firstly, we assume that P*S* can crossphosphorylate targets within the spatially proximal PLK4-STIL complexes. Secondly, we posit that the PLK4-STIL complexes phosphorylated on STIL, PS* and P*S*, possess longer centriolar residence time than PLK4 itself and PLK4-STIL complexes not phosphorylated on STIL. Therefore, SAS6, which interacts only with phosphorylated STIL and thus promotes centriolar retention of P*S* and PS*, is included in our model implicitly.

We model centriole replication as an explicitly open system: both PLK4 and STIL are continuously produced throughout procentriole biogenesis, while the complexes of phosphorylated PLK4, P*S and P*S*, are subject to degradation. We estimated the rates of P*S and P*S* degradation based on the half-life time of PLK4 experimentally measured to be 2h (Klebba et al., 2015). The spatial domain of our model is represented by a cylindrical shell immersed into a homogenous cytoplasm. The cylinder has dimensions characteristic of the proximal end of a mammalian centriole and is subdivided into $N = 9$ identical vertical stripes, distinct compartments within which all molecular concentrations are deemed spatially uniform. Detailed model equations and simulation parameters are provided in the Supplemental Information.

## The model predicts robust PLK4 symmetry breaking from ring to spot

We first simulate cellular dynamics of PLK4 in G1 phase by assuming that PLK4, absent at the simulation start, begins to accumulate at a slow constant rate. Cytoplasmic PLK4 then reversibly binds to the centriole and equally populates all nine compartments producing the characteristic symmetric ring pattern of PLK4 localization (Figures 2A,B and S1B). With a two-hour delay (arrowhead on Figure 2B), STIL also starts to express at a constant rate. For approximately one hour both proteins progressively accumulate on the centriole, equally in all compartments. At ca. 3h past the start of PLK4 expression (arrow on Figure 2B), this spatially uniform regime exhibits a dramatic instability during which every spatial compartment behaves differently from others. The simulation shown in Figure 2A exhibits the characteristic features of this symmetry-breaking transition. In under five minutes, a uniform ring that existed for nearly three hours (first frame) is replaced by an asymmetric distribution with two distinct maxima separated by the compartments with rapidly vanishing PLK4. The two maxima grow together for 20 min, but with slightly different rates. From 3:24 (third frame), only the dominant maximum continues to grow while the other declines. Finally, a unique spot of PLK4 is established by 4h and remains stable thereafter symbolizing the emergence of a single daughter centriole. This example demonstrates that the transition from ring to spot may involve intermediate short-lived states with multiple maxima of PLK4 localization and activity. Extensive variation of model parameters reveals that a single procentriole is robustly produced in a wide range of parameters. Nevertheless, essentially each model parameter can be altered so that multiple procentrioles are produced with the parameter-dependent probabilities. We defer discussion of this observation until the last section where we explore how the number of procentrioles changes as the protein production and degradation rates are varied simultaneously.

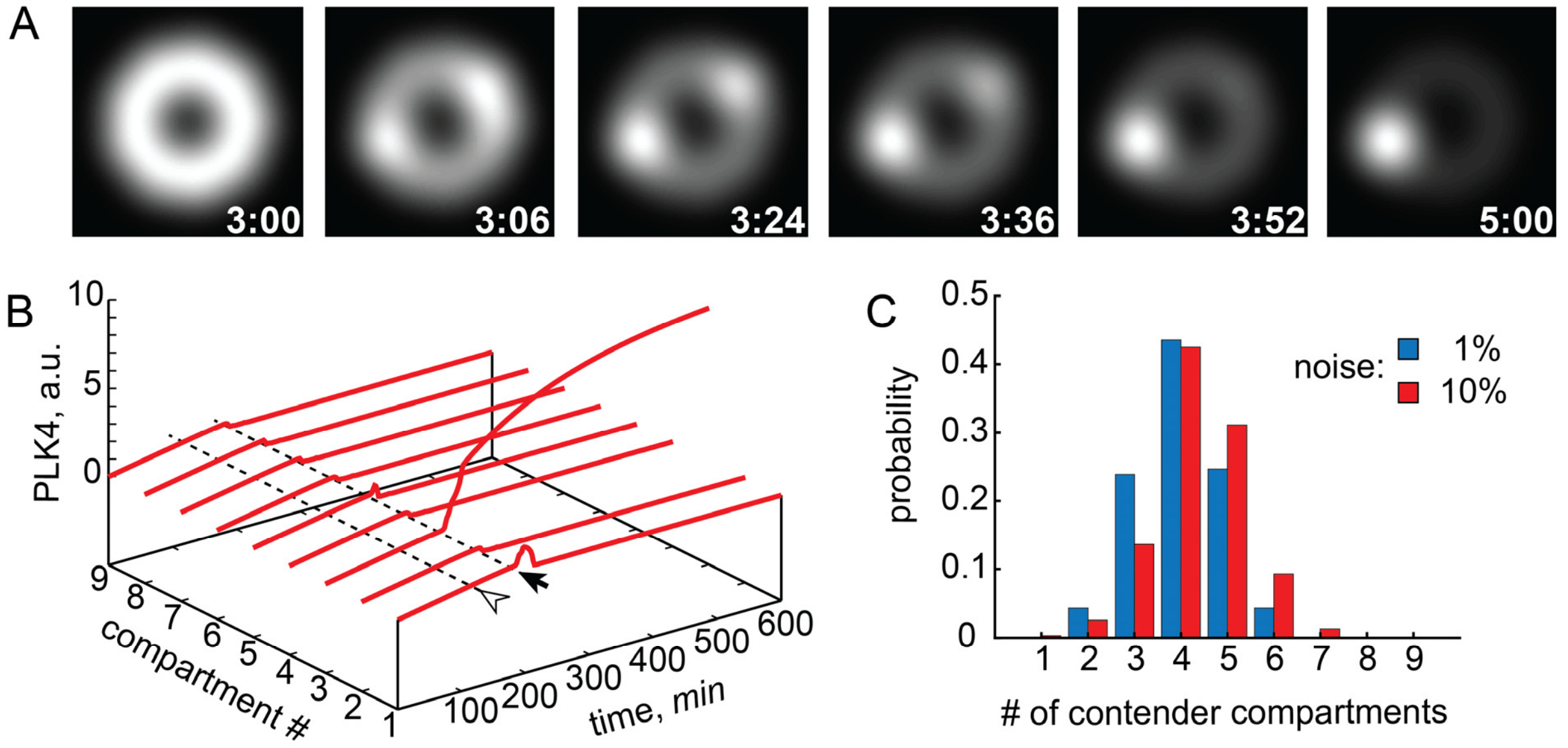

**Figure 2. Stochastic model of procentriole formation exhibits spontaneous symmetry breaking of PLK4 localization from "ring" to "spot".** (A) A simulation showing symmetry breaking scenario via a prolonged coexistence of two PLK4 maxima. PLK4 surrounding the centriole is shown as simulated fluorescence stills at the indicated time points (hr:min). (B) Symmetry breaking in a distinct stochastic realization of the model with the same parameters as in A. Total quantity of PLK4 in all compartments is shown as a time series. Arrowhead indicates the time point at which STIL begins to express. Arrow points to the onset of instability of the symmetric localization of PLK4. (C) Probability distribution of the number of contender compartments which attempt to increase their PLK4 content during the symmetry breaking. Histograms are computed at two shown levels of molecular noise, 400 simulations each. See also Figures S1 and S2.

The questions of how and why a single locus on the surface of mother centriole appears to be chosen to build a procentriole are frequently raised in the literature. To address them, we performed extensive stochastic simulations in which we scored each compartment as a contender if it "attempted" to build a procentriole, or a non-contender otherwise. To qualify as a contender, a compartment had to have a PLK4 level greater than that of the spatially uniform state for at least 6 min, the lower limit for the duration of the symmetry-breaking transition as observed in our simulations. Surprisingly, our results show that with both high and low levels of simulated molecular noise, on average half of compartments attempt to increase their PLK4 level (Figure 2C). Notwithstanding, two daughter centrioles were formed instead of one in only 3 out of 400 simulations (0.75%). Thus, in the presence of inevitable molecular noise each locus has a 50% chance of increasing its PLK4 level and, consequently, equal initial potential to form a procentriole. We conclude that formation of a single procentriole is not a result of passive memorization of a random site that was chosen early in the process of procentriole formation. Instead, the existence of multiple contender compartments in our model suggests the existence of active process(es) that are responsible for the selection of only one among them.

## PLK4 autocatalysis, degradation, and activity-dependent retention of PLK4-STIL complexes are necessary for symmetry breaking

We next sought to determine which biochemical reactions that comprise the network presented in Figure 1 are essential for the breaking of symmetry. We first checked that changing the number of centriole spatial compartments, $N$, does not qualitatively affect the behavior of the model. To keep mathematical analysis tractable (see "Stability analysis of stationary states" in the Supplemental Information), we then reduced the number of centriolar compartments to $N = 2$, which is sufficient to observe symmetry breaking. First, we varied the rates of protein expression and degradation. To reduce the dimensionality of the analysis, respective rates for PLK4 and STIL were kept equal. The results shown in Figure 3A demonstrate that degradation is indeed required for symmetry breaking. At a fixed level of protein expression, there exists a threshold degradation rate below which symmetric ring-shaped localization of overexpressed PLK4 remains stable. On the opposite end of the interval of symmetry breaking, the model predicts a maximal level of degradation above which a new symmetric state, now with very little PLK4 associated with the centriole, is found. Reciprocally, at a fixed rate of degradation, both increasing the expression past a certain maximal level and decreasing it below the threshold again results in a stable PLK4 ring. The model thus faithfully recapitulates the results of experiments with PLK4 amplification by overexpression and expression of a non-degradable mutant, as well as PLK4 reduction by both slower production (siRNA) and faster degradation (Bettencourt-Dias et al., 2005; Habedanck et al., 2005; Kleylein-Sohn et al., 2007; Lambrus et al., 2015; Rogers et al., 2009). We conclude that the rates of PLK4 expression and degradation must be carefully balanced to enable centriole duplication.

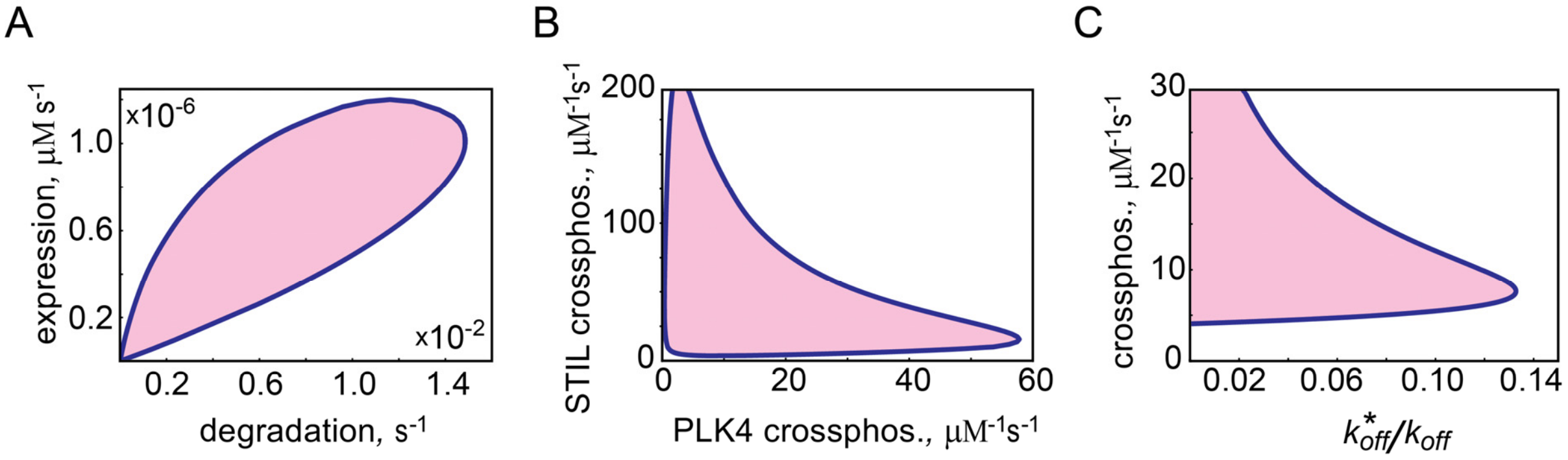

**Figure 3. PLK4 symmetry breaking requires a balance of protein expression and degradation, and positive feedback based on the PLK4 kinase activity.** Domain of symmetry breaking is shown by color in all panels. (A) Symmetry breaking occurs at the optimal combination of protein expression and degradation rates. Respective PLK4 and STIL parameters were kept equal. (B) Symmetry breaking requires autocatalytic crossphosphorylation of both PLK4 and STIL. (C) Crossphosphorylation and retention of phosphorylated PLK4-STIL complexes are the two parts of the dual PLK4 activity-based feedback required for procentriole formation. *x* axis represents the ratio of the off rate $k^*_{off}$ for the PLK4-STIL complexes phosphorylated on STIL (PS*, P*S*) to the off rate $k_{off}$ of P, PS and P*S. Crossphosphorylation rates of STIL and PLK4 were kept equal.

Crossphosphorylation postulated in our model represents a type of autocatalytic amplification in which more molecules of fully phosphorylated PLK4-STIL complex P*S* are produced by P*S* from the less phosphorylated complexes:

$$\mathrm{PS} + \mathrm{P^*S^*} \rightarrow \mathrm{P^*S} + \mathrm{P^*S^*} \rightarrow 2\mathrm{P^*S^*},$$
$$\mathrm{PS} + \mathrm{P^*S^*} \rightarrow \mathrm{PS^*} + \mathrm{P^*S^*} \rightarrow 2\mathrm{P^*S^*}.$$

Biochemically, crossphosphorylation reactions fall into two distinct classes in which PLK4 phosphorylates either another PLK4 or STIL molecule within a distinct complex. Interestingly, our model shows that both types of crossphosphorylation are required together for symmetry breaking, while neither can compensate for the complete absence of the other. This can be seen in Figure 3B where the zone of symmetry breaking does not touch either axis. The model thus predicts that a weak crossphosphorylation activity of one type can be compensated by the elevated activity of the other type.

In formulating our model, we proposed that the initiation of centriole duplication is induced by a dual positive feedback based on the PLK4 kinase activity. We hypothesized that autocatalytic crossphosphorylation increases both the local activity and the local concentration of PLK4. The increase in surface density is achieved in our model by decreasing the off rate $k^*_{off}$ of PLK4-STIL complexes phosphorylated on STIL, PS* and P*S*, thus increasing their centriole retention. Our results demonstrate that the ratio of the off rates $k^*_{off}/k_{off}$ , where $k_{off}$ denotes the off rates of P, PS and P*S, may not exceed a certain maximal value, regardless of the strength of autocatalysis (Figure 3C). At the same time, even a very large difference in the off rates ($k^*_{off}/k_{off} \ll 1$) cannot compensate for the lack of autocatalysis. Our results are thus fully consistent with the experiments in which application of the PLK4 kinase activity inhibitor, centrinone, resulted in the failure to break symmetry and duplicate the centrioles (Ohta et al., 2018; Wong et al., 2015). As in experiments of Ohta *et al.*, inhibition of PLK4 kinase activity produces in our model progressive accumulation of inactive PLK4 in the form of a symmetric ring surrounding the mother centriole. We conclude that the dual positive feedback based on the activity of the PLK4 kinase destabilizes the spatially symmetric distribution of PLK4-STIL complexes and induces self-organization of nascent procentriole.

**Single procentriole emerges from the competition for PLK4 and STIL**

Our results show that procentriole initiation starts in multiple spatial loci simultaneously and, therefore, the final emergence of a unique procentriole is not simply the consequence of one site being randomly selected from the outset. While the duration of coexistence is particularly prolonged in Figure 2A (50 min), essentially all simulations exhibit a short-lived presence of at least one extra PLK4 maximum, in addition to the one destined to become “the spot” (see, e.g., Figure 2B with two such maxima). The dynamics with which multiple PLK4 maxima resolve into a single spot, suggests that these maxima compete for a common resource. To identify this resource, we performed a detailed analysis of simulations in which two contender compartments initially exhibit rapid accumulation of PLK4-STIL complexes but then, with a slower kinetics, one of the two compartments loses its protein content and the other becomes the procentriole (see Figure 4 for a representative example of such a simulation).

Which process is responsible for the differential fate of the two initially successful contenders? We hypothesized that this outcome is mediated by the exchange of proteins between the mother centriole surface and the cytoplasm. To test this hypothesis explicitly we calculated the centriole-cytoplasmic flux of PLK4 in all nine spatial compartments. Figure 4B shows that, at symmetry breaking, both contenders exhibit rapid intake of PLK4 (positive flux, red and blue lines), while non-contender compartments release PLK4 (negative flux, green line). After this initial peak, the behavior in the two contender compartments is distinctly different. One continues to accumulate PLK4 as demonstrated by a slowly increasing positive flux, while the PLK4 flux in the other begins to diminish and eventually becomes negative (Figure 4B). From the moment when the PLK4 flux changes sign, the unsuccessful contender releases its PLK4 content back to the cytoplasm and this release, rather than degradation, is responsible for the rapid disappearance of the protein content in the unsuccessful contender.

We next asked if this recycled PLK4 contributes to the PLK4 increase seen in the winning compartment. To address this question, we performed an in-silico “photoactivation” (PA) of PLK4 in the losing contender compartment. Namely, all PLK4 molecules residing within this compartment were virtually labeled at the time point indicated by vertical line in Figure 4A. Figure 4C demonstrates that a fraction of the PA PLK4 released by the losing contender into the cytoplasm was re-adsorbed back by the centriole and the winning compartment got the most of this PA PLK4. We conclude that competition between the intermediate maxima of PLK4 is achieved via the cytoplasm-mediated exchange of PLK4-STIL complexes. This conclusion is non-trivial because the system is not mass-conserved, and the proteins are continuously synthesized and degraded. However, the characteristic time of the transition from the PLK4 ring to spot is much shorter than 2 h, the experimentally determined half-life time of PLK4 (Klebba et al., 2015). Indeed, at the chosen model parameters, the duration of coexistence of the two largest intermediate maxima takes on average 13 min but, frequently, a single spot establishes within only 6-10 minutes. Thus, during the ~10 min time interval within which symmetry breaking occurs, the change in the total cellular PLK4 and STIL due to their expression and degradation is very small. Given that both ring and spot exist in the model for hours, such a rapid transition from ring to spot might explain why intermediate states between the ring and the spot are difficult to visualize in experiments imaging endogenous PLK4 (Ohta et al., 2014; Ohta et al., 2018). Nevertheless, temporally tracking the PLK4 ring to spot transition in *Drosophila* has revealed intermediate states with multiple PLK4 maxima, which could correspond to the model contender sites (Dzhindzhev et al., 2017). Possibly, they had been also observed in experiments with PLK4 overexpression as a "halo" surrounding mother centrioles (Kleylein-Sohn et al., 2007). We thus propose that the coexistence between spatial loci on the surface of centriole becomes spontaneously unstable at some threshold level of PLK4 accumulation and its activation by STIL. Instead, the loci engage into an antagonistic winner-takes-all competition for the PLK4-STIL complexes. Under physiologically normal intracellular conditions this competition resolves in the formation of only one daughter centriole.

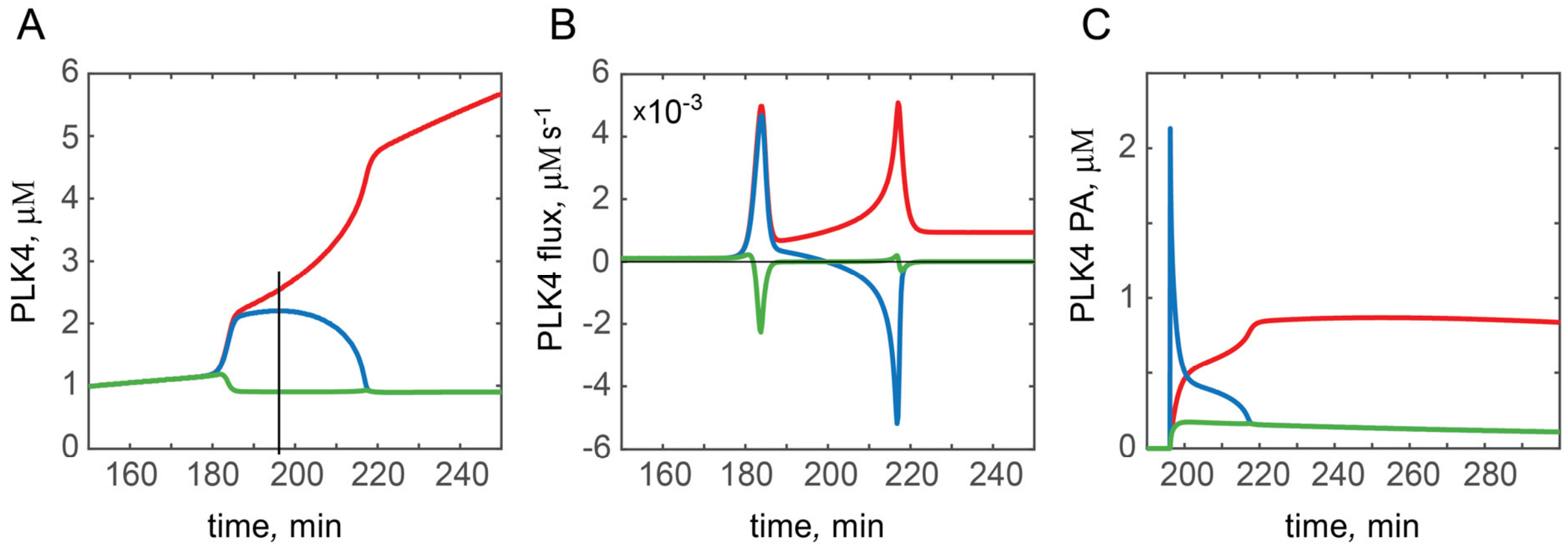

**Figure 4. A single procentriole is established by the competition for PLK4 and STIL**. Temporary dynamics in the winning (red), unsuccessful contender (blue), and a typical non-contender (green) model compartments is shown for one representative simulation. (A) Total PLK4. Vertical line indicates time of the *in-silico* photoactivation of PLK4 in the unsuccessful contender compartment. (B) Centriole-cytoplasmic flux of PLK4. (C) Dynamics of the *in-silico* photoactivated PLK4. See also Figure S2.

## Degradation without competition does not break the symmetry

Could an alternative molecular mechanism, not involving competition via the cytoplasmic exchange of proteins, explain the formation of a unique procentriole? Indeed, in situ degradation of PLK4 and its regulation by STIL had been proposed to explain procentriole formation (Arquint et al., 2015; Ohta et al., 2014; Ohta et al., 2018). To test this hypothesis in the model, we first abrogated competition between the centriole spatial compartments. Since competition is achieved by exchanging proteins via the common cytoplasm, we prevented this exchange between the surface of centriole and the cytoplasm by reducing the off rates for PLK4, STIL, and all their complexes (PS, P*S, PS*, P*S*) to zero. This implies that once a molecule of PLK4 or STILL is bound to the centriole it can undergo biochemical transformations and degradation *in situ*, but it may not leave the surface of the centriole.

We first asked whether the model can still generate symmetry breaking when competition between compartments is abolished. To allow for rigorous mathematical analysis, we again resorted to the case with $N = 2$ centriolar compartments. The results of this analysis (see Supplemental Information) demonstrate that disruption of protein recycling back to the cytoplasm prevents symmetry breaking. Qualitative diagrams shown in Figure 5 compare the behavior of the model with and without competition. Temporary dynamics of the model in the multidimensional space of its variables is routinely represented by a trajectory directed towards one of the stable steady states. Figure 5 qualitatively shows the dynamics of model with $N = 2$ centriolar compartments. Only the behavior of the autocatalytic PLK4-STIL complex P*S* in both compartments is shown to reduce the dimensionality of presentation. In the first scenario, competition between the two compartments is prevented by abrogation of protein recycling (Figure 5A). As PLK4 and STIL gradually accumulate in the model, the trajectories start at the origin (0,0) and invariably arrive at the only stable steady state. Since this symmetric state is globally stable in the absence of competition, molecular noise cannot destabilize it regardless of the amplitude. A qualitatively different behavior is observed in the second scenario, where exchange of proteins via the common cytoplasm is enabled (Figure 5B). Autocatalytic amplification of PLK4 activity in the presence of protein exchange destabilizes the coexistence between compartments and the symmetric state of the centriole becomes an unstable steady state of the saddle type. Trajectories started at the origin are still attracted to this state, but in its close vicinity they deflect towards one of the two stable asymmetric states (purple arrows in Figure 5B). Even a small-amplitude molecular noise can drive symmetry breaking and decide which of the two states is chosen by the system.

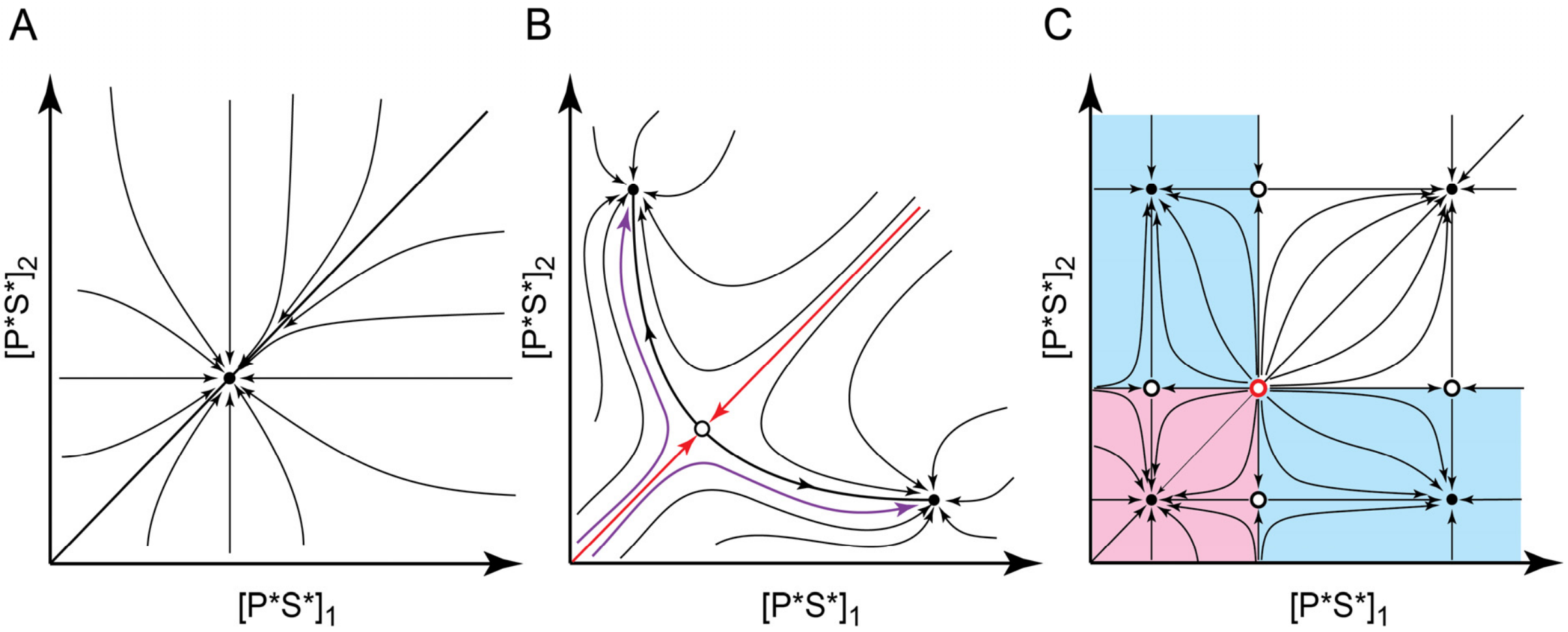

**Figure 5. Degradation without competition cannot break PLK4 symmetry.** Phase space dynamics of the model with two centriolar compartments is shown qualitatively for three different scenarios: (A) Linear degradation of PLK4 without competition between the compartments; (B) Linear degradation of PLK4 with competition (base model); (C) Nonlinear degradation of PLK4 without competition. Arrows indicate direction of temporary dynamics. Stable steady states are shown as filled circles, unstable steady states are denoted by open circles (saddles, black; a repeller, red). In (B): red arrows indicate trajectories separating basins of attraction of two stable states, purple arrows show typical system trajectories in the presence of molecular noise. In (C): basins of attraction are shown in color (low symmetric state, light magenta; asymmetric states, light cyan).

We next considered the possibility that our model cannot explain symmetry breaking in the absence of competition because its approach to degradation is oversimplified. Until now we assumed that the degradation rate of complexes of phosphorylated PLK4 is simply proportional to their concentration. It had been proposed, however, that the binding of STIL somehow protects PLK4 from degradation *in situ* (Arquint et al., 2015; Arquint and Nigg, 2016; Ohta et al., 2014; Ohta et al., 2018). How could such a protection effect be reconciled with STIL promoting the kinase activity of PLK4 and, therefore, its degradation? One possibility is suggested by the recent results that show that SMN, another target of the PLK4 ubiquitin E3 ligase $SCF^{Slimb}$, avoids degradation at high spatial density by sequestering its phosphodegron within high-order multimeric complexes (Gray et al., 2018). Therefore, STIL could directly promote degradation of PLK4 by increasing its kinase activity and indirectly protect PLK4 by driving formation of spatially dense PLK4-STIL complexes within which the interaction of the PLK4 phosphodegron with $SCF^{Slimb}$ is diminished. To translate this hypothesis into the model, we assume that *in situ* degradation of PLK4-STIL complexes P*S* and P*S, in addition to a weak linear term, is also described by a term that vanishes at their high spatial density, i.e.,

$$X_{\deg} = -a_1 X - \frac{a_2 X}{X^2 + a_3^2},\ X = [P^*S^*],[P^*S]\ ,$$

where $a_1, a_2, a_3$ are constants. In this mathematical formulation degradation of P*S and P*S* is inhibited at their high spatial density (see Supplemental Information). Combined with continuous influx of PLK4 and STIL from the cytoplasm, this additional assumption of nonlinear degradation converts each spatial compartment into a bistable system. At the same cytosolic concentrations of PLK4 and STIL, a compartment can be present in two distinct stable states with either low or high density of PLK4-STIL complexes. The centriole could then hypothetically exhibit asymmetric configurations, e.g., with only one compartment in the high PLK4 state, even in the absence of protein recycling and, thus, competition between the compartments. Surprisingly, however, simulations in which PLK4 and STIL cellular content gradually increases, invariably produce only the symmetric state with all spatial compartments in the low PLK4 state, regardless of the number of compartments used or the magnitude of molecular noise.

The interpretation of this result can be aided by Figure 5C that qualitatively illustrates behavior of the system with nonlinear degradation. Here each of the two compartments can be present in two stable states independently of the state of the other compartment. These states are separated by saddles whose positions determine which of the stable states has a larger basin of attraction. System trajectories that start at the origin (0,0) invariably lie within the basin of attraction of the lower symmetric state (magenta domain in Figure 5C). Although formally this state is stable only locally, in practice, it would require improbably high molecular noise to force the system out of this stable state into one of the basins of attraction for the asymmetric states (cyan domains). Note that by changing model parameters, it is possible to move the separating saddle arbitrarily close to the symmetric low state and, thus, destabilize it. However, this parameter change also destabilizes both asymmetric states to the same extent. As a result, molecular noise would push the system not into one of the asymmetric states but, instead, into the symmetric high state, failing to break symmetry of the centriole. Thus, surprisingly, this model with nonlinear degradation, is essentially as insensitive to noise as the one with linear degradation but no competition (Figure 5A). We conclude that the introduction of nonlinear degradation does not rescue symmetry breaking in the system where competition between spatial compartments is prevented.

**Overexpression of PLK4 and STIL produces supernumerary procentrioles**

Experiments with overexpression of PLK4, STIL and other core proteins required for centriole duplication led to supernumerary procentrioles arranged around mother centriole in a characteristic rosette pattern. We asked if our model can generate supernumerary centrioles and reproduce the characteristic quantitative traits observed in overexpression experiments. To reduce the dimensionality of the analysis, we assumed that PLK4 and STIL are overexpressed equally, in a 1:1 stoichiometric ratio, and performed numeric analysis of our stochastic model with $N = 9$ spatial compartments. We adopted the model parameter set that was used to produce the results shown in Figure 2 as our baseline since this set of parameters generates a single procentriole with very high fidelity (over 99% of trials produce a single procentriole). We found then that the tenfold increase in the rate of protein production (henceforth, overexpression for brevity) resulted in the loss of symmetry breaking. We thus set out to explore the outcome of simulations with intermediate overexpression factors ranging between 1 and 10.

Overexpression of PLK4 and STIL revealed two major traits in the model behavior (Figure 6A). Firstly, as the overexpression factor increments, the model produces a progressively increasing number $n = 1, 2, ...8$ of equal PLK4 maxima representing the emergence of $n$ identical procentrioles. Importantly, despite the ongoing competition between the PLK4 maxima, these multiple procentrioles are stable steady states of our model. We conclude that overexpression stabilizes multiple procentrioles which are unstable under the normal physiological rates of expression. Secondly, as protein overexpression increases, both the most likely number and variability in the number of produced procentrioles grows. Thus, at x2 overexpression, ~93% of 400 simulations generate two procentrioles (Figure 6A). The remaining 7% of simulations produced exactly 3 procentrioles, while patterns with more than three procentrioles were not found. At x6 overexpression, however, the most likely number of procentrioles is 5, while 4, 6 and even 7 procentrioles were identified among the outcomes of simulations (Figure 6A). These results suggest the existence of a sliding window of probability that determines which numbers of procentrioles can be observed with the given model parameters. Both the position and width of this window increase with overexpression. However, the change in width is much less pronounced than the change in the position as even at x8 overexpression some numbers of procentrioles ($n = 1, 2, 3, 4$) cannot be realized. Remarkably, a very similar quantitative trend had been observed in experiments where the number of centrioles had been carefully assayed in response to progressive increase in PLK4 abundance (Kleylein-Sohn et al., 2007).

Finally, we sought to explore how various patterns of supernumerary procentrioles are distributed on the 2D plane of rates of protein expression and degradation. Towards this goal we computed the most likely number of procentrioles on a rectangular grid of chosen parameters as shown by color in Figure 6B. This systematic variation of parameters confirms our observation that the most likely number of procentrioles produced per mother centriole increases progressively with protein overexpression. The domains of parameters corresponding to distinct $n$ have comparable widths that only slightly decrease with $n$. The apparent fuzzy appearance of the boundaries between these domains reflects the stochastic nature of the model and shows that the switch between the dominant patterns is achieved via a gradual change in their probabilities (Figure 6B). Note that while the rates of expression and degradation are natural parameters to vary to induce multiple procentrioles, specific changes in other model parameters can produce the same outcome. Thus, the model predicts that altering biochemical parameters other than those that control the abundance of PLK4 and STIL can also affect the number of procentrioles formed. Taken together with experimental observations, our results highlight a remarkable property of the numerical control of centriole replication. Under the physiologically normal system parameters (e.g., the rates of protein expression, degradation and biochemical reactions), the outcome of symmetry breaking is highly robust and insensitive to the molecular noise. A single procentriole is produced with exceptional fidelity and overduplication is essentially non-existent.

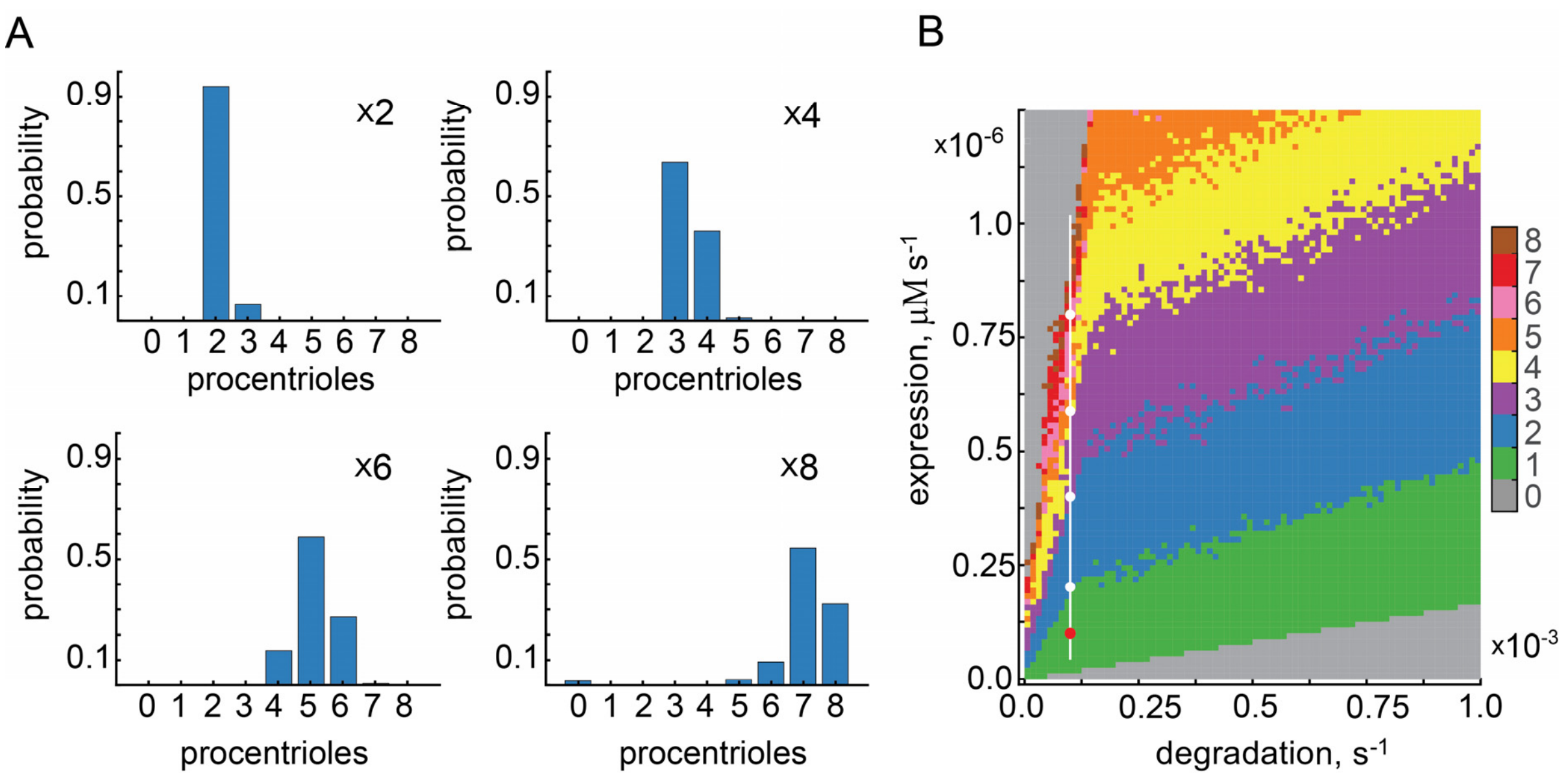

**Figure 6. Levels of expression and degradation of PLK4 and STIL control the number of procentrioles formed during symmetry breaking.** (A) Protein overexpression produces supernumerary procentrioles. Overexpression factor is shown on the panels. Histograms were produced based on 400 simulations. (B) The most likely number of procentrioles as a function of rates of expression and degradation. Ten simulations were performed at each grid point. The baseline set of model parameters corresponding to a single procentriole is indicated by the red dot. Overexpression conditions used in (A) are shown by white dots.

## Conclusions

We propose a realistic biophysical model that explains both symmetry breaking of the spatially uniform distribution of PLK4 around the mother centriole (“ring”) and formation of the unique cluster of PLK4-STIL complexes (“spot”) that initiates biogenesis of the procentriole. Positive feedback that drives symmetry breaking consists of two converging arms (Figure 7). In the first arm, the autocatalytic crossphosphorylation of PLK4, provides local autoamplification of PLK4 activity in the presence of opposing phosphatase(s). Such density-dependent activation appears to be a common property among mitotic kinases, such as Aurora B (Zaytsev et al., 2016). The second arm is mediated by the activator-scaffold STIL and provides activity-dependent retention of active PLK4 on the surface of centriole.

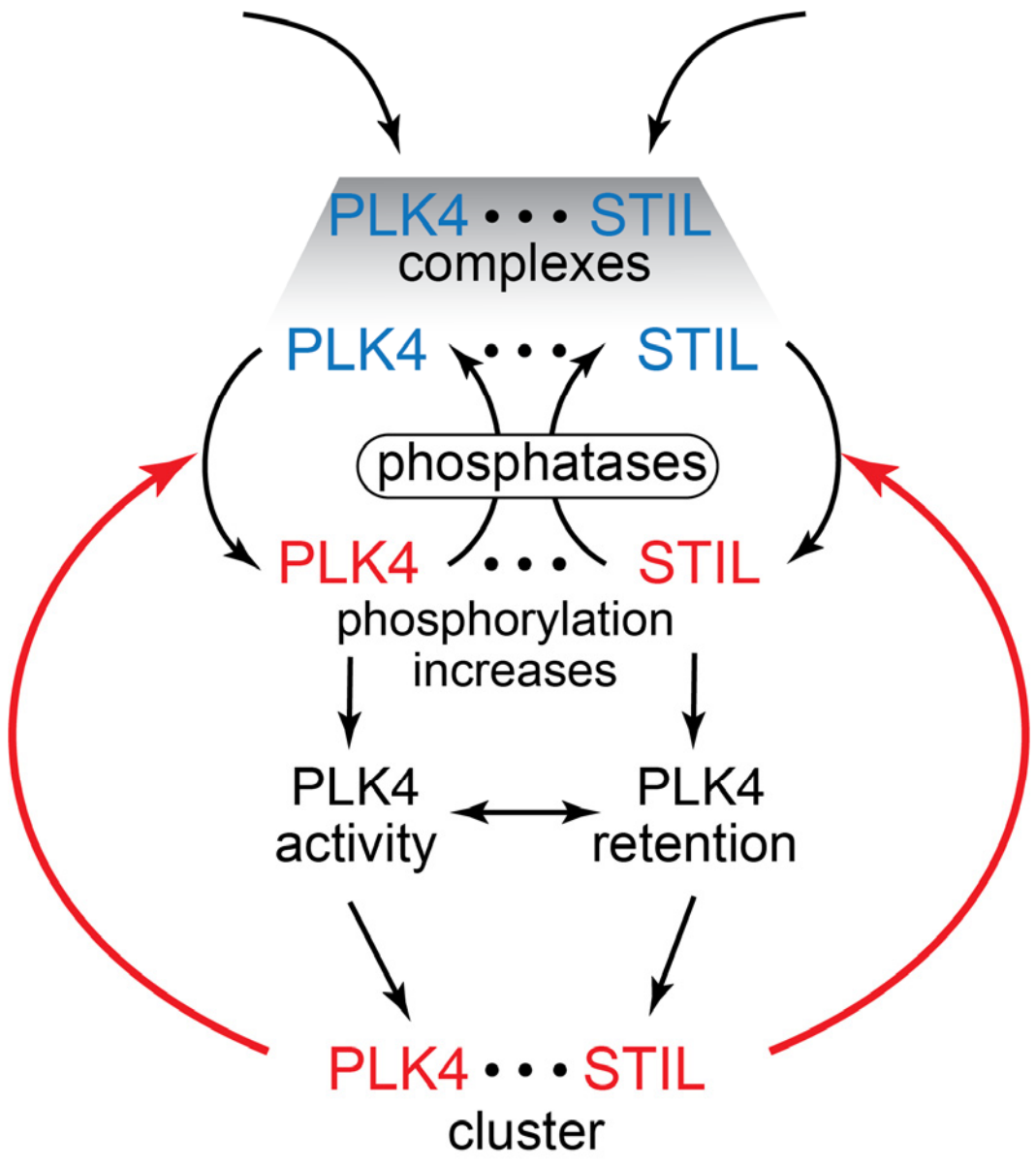

**Figure 7. Dual positive feedback drives symmetry breaking of PLK4-STIL complexes on the surface of centriole.** Phosphorylated molecules are shown in red, unphosphorylated in blue. Feedback loops are shown by red arrows. Horizontal dots ••• symbolize bonds between PLK4 and STIL.

While active PLK4 cannot directly "recruit" inactive cytoplasmic PLK4, activity-dependent retention can readily provide localized accumulation of active PLK4 even if inactive cytoplasmic PLK4 binds to the centriole spatially uniformly. Thus, this mechanism provides a robust biophysical explanation to the proposed in the literature self-recruitment of PLK4 during the procentriole formation (Aydogan et al., 2018). Interestingly, albeit the molecular mechanism is different, PLK1 and possibly other PLK-family kinases also exhibit activity-dependent retention (Park et al., 2011).

An important feature of our model is that the two feedback arms synergize (Figure 7). Indeed, activity-dependent retention of PLK4-STIL complexes on the surface of the centriole promotes spatial clustering and, thus, crossphosphorylation. In turn, crossphosphorylation enhances the retention of PLK4-STIL complexes. As larger protein complexes have smaller diffusive mobility in the cytoplasm, this dual positive feedback could also potentially provide symmetry breaking in the initially spatially homogeneous cytoplasm, away from any pre-existing centrioles, and, thus, also explain the *de novo* centriole formation. An important consequence of the activity-dependent retention is that out of two unequal PLK4-STIL clusters, the bigger one will grow faster. This property undermines neutral coexistence of multiple PLK4-STIL clusters. Because the clusters can exchange their material via the common cytoplasm, they, in fact, engage in an antagonistic winner-takes-all competition, which, under physiologically normal conditions, results in the emergence of a single procentriole. We, therefore, predict that improvement in the spatial and temporal resolution of live-cell imaging of centriole duplication will reveal emergence of multiple competing PLK4 maxima and their subsequent resolution towards a single procentriole. We hypothesize that the principles highlighted by our model are instrumental for self-organization of unique cellular structures, regardless of the details of the molecular mechanisms. While our model is formulated for mammalian cells, the principles of PLK4 symmetry breaking outlined above are likely to be conserved also in other organisms. Further concerted experimental and theoretical efforts will be required to increase the biological realism and predictive power of the model by the refinement of molecular mechanisms and incorporation of additional molecular players, such as CEP85, which was recently implicated in PLK4 activation (Liu et al., 2018).

## Supplemental information

Supplemental Information includes Methods, two figures and two tables.

## Acknowledgements

ABG acknowledges financial support from the Biotechnology and Biological Sciences Research Council of UK (BB/P01190X, BB/P006507). Work in the AJH laboratory was supported by a R01 research grant from the National Institutes of Health (GM 114119), an American Cancer Society Scholar Grant (129742-RSG-16-156-01-CCG), and a grant from the Pew-Steward Trust.

## Author contributions

M.L., A.J.H., and A.B.G. designed the study. M.L. and A.B.G. developed and analyzed the mathematical models. M.L., A.J.H., and A.B.G. wrote and revised the paper.

## Declaration of interests

The authors declare no competing interests.

**Supplemental Information**

**Mathematical formulation of the model**

We estimated that the proximal end of a centriole is a cylinder with radius *R = 250nm* and height *h = 400nm* (Lawo et al., 2012; Sonnen et al., 2012) that is immersed into the spatially homogeneous cytoplasm. Rather than assuming that PLK4 and STIL bind to and react on a 2D surface, we postulated that these reactions happen within a volume determined by a cylindrical shell with finite thickness. This choice is motivated by the fact that binding sites for PLK4 are molecularly determined by the N-termini of CEP152 and CEP192, which are thought to be positioned 200-220*nm* and ~300*nm* away from the center of the centriole, respectively (Park et al., 2014; Sonnen et al., 2013; Sonnen et al., 2012). For simplicity, we assumed that these fixed binding sites are approximately uniformly distributed within the reaction volume. We then subdivided it into *N* = 2 or 9 equal compartments (Figure S1B). Therefore, within each compartment, numbered by index *i = 1,2…N,* the concentrations of free PLK4 ([$P_i$]) and PLK4-STIL complexes ([$PS_i$], [$PS^*_i$], [$P^*S_i$], and [$P^*S^*_i$]) were assumed spatially homogeneous. Compartments are connected by the common cytoplasm where the concentrations of free PLK4 ([$P_c$]), free STIL ([$S_c$]) and their complexes ([$PS_c$], [$PS^*_c$], [$P^*S_c$] and [$P^*S^*_c$]) are also spatially homogeneous.

The following six types of reactions take place in every centriole compartment and/or in the cytoplasm:

1) Phosphorylation of PLK4 and STIL within the same PLK4-STIL complex. The corresponding reaction rates in the cytoplasm ($R_{1,c}$, $R_{2,c}$, $R_{3,c}$ and $R_{4,c}$) and on the centriole ($R_{1,i}$, $R_{2,i}$, $R_{3,i}$ and $R_{4,i}$) are given as follows:
$$R_{1,c} = k_1[PS_c], R_{2,c} = k_2[PS_c], R_{3,c} = k_3[PS^*{}_c], R_{4,c} = k_4[P^*S_c]$$
$$R_{1,i} = k_1[PS_i], R_{2,i} = k_2[PS_i], R_{3,i} = k_3[PS^*{}_i], R_{4,i} = k_4[P^*S_i]$$

2) Dephosphorylation of PLK4 and STIL with the reaction rates in the cytoplasm ($R_{-1,c}$, $R_{-2,c}$, $R_{-3,c}$ and $R_{-4,c}$) and on the centriole ($R_{-1,i}$, $R_{-2,i}$, $R_{-3,i}$ and $R_{-4,i}$):
$$R_{-1,c} = k_{-1}[PS^*{}_c],\ R_{-2,c} = k_{-2}[P^*S_c],\ R_{-3,c} = k_{-3}[P^*S^*{}_c],\ R_{-4,c} = k_{-4}[P^*S^*{}_c]$$
$$R_{-1,i} = k_{-1}[PS^*{}_i],\ R_{-2,i} = k_{-2}[P^*S_i],\ R_{-3,i} = k_{-3}[P^*S^*{}_i],\ R_{-4,i} = k_{-4}[P^*S^*{}_i]$$

3) Crossphosphorylation of PLK4 and STIL on the centriole ($R_{5,i}$, $R_{6,i}$, $R_{7,i}$ and $R_{8,i}$).
$$R_{5,i} = k_5[PS^*{}_i][P^*S^*{}_i], R_{6,i} = k_6[P^*S_i][P^*S^*{}_i], R_{7,i} = k_7[PS_i][P^*S^*{}_i],$$
$$R_{8,i} = k_8[PS_i][P^*S^*{}_i]$$

4) Binding/unbinding of free PLK4 and PLK4-STIL complexes from the cytoplasm to the centriole ($R_{10,i}$, $R_{11,i}$, $R_{12,i}$ , $R_{14,i}$, $R_{15,i}$ and $R_{16,i}$). Here and in 5) parameters *a* and *b* represent concentration conversion factors arising because molecules move between centriolar compartments and the cytoplasm. Precise values of these parameters depend on the specific reaction and are given below in the formulation of the model equations.
$$R_{10,i}(a,b) = k_{10}[PS_c]a - k_{-10}[PS_i]b,$$
$$R_{11,i}(a,b) = k_{11}[PS^*{}_c]a - k_{-11}[PS^*{}_i]b,$$
$$R_{12,i}(a,b) = k_{12}[P^*S_c]a - k_{-12}[P^*S_i]b,$$
$$R_{14,i}(a,b) = k_{14}[P^*S^*{}_c]a - k_{-14}[P^*S^*{}_i]b,$$
$$R_{16,i}(a,b) = k_{16}[P_c]a - k_{-16}[P_i]b$$

5) Association/dissociation of free PLK4 and STIL in the cytoplasm ($R_{15,c}$) or on the centriole ($R_{15,i}$).

$$R_{15,c} = k_{15}[P_c][S_c] - k_{-15}[PS_c],$$
$$R_{15,i}(a,b) = k_{15}[P_i][S_c]a * N_i - k_{-15}[PS_i]b,$$

here $N_i = 1 + am * (U(0,1) - 0.5)$ is the molecular noise term, U(0,1) is the uniform distribution between 0 and 1 produced by a standard random number generator, *am < 1* is the noise amplitude.

6) Production of the free PLK4 ($P_c$) and STIL ($S_c$) in the cytoplasm ($R_{24}$ and $R_{25}$, respectively).

$$R_{24} = k_{24}, R_{25} = 0 \; if (t < t_{STIL}), \; otherwise \; R_{25} = k_{25}$$

7) Protein degradation. Free unphosphorylated PLK4 ($P_c$) and STIL ($S_c$) are assumed to slowly degrade in the cytoplasm with reaction rates $R_{-24}$ and $R_{-25}$, respectively. PLK4-STIL complexes in which PLK4 is phosphorylated (P*S and P*S*) are degraded in the cytoplasm ($R_{26,c}$ and $R_{27,c}$) and *in situ* on the centriole ($R_{26,i}$ and $R_{27,i}$):

$$R_{-24} = k_{-24}[P_c], R_{-25} = k_{-25}[S_c]$$
$$R_{26,c} = k_{26,c}[P^*S_c], R_{27,c} = k_{27,c}[P^*S^*{}_c]$$
$$R_{26,i} = k_{26}[P^*S_i], R_{27,i} = k_{27}[P^*S^*{}_i]$$

7a) In the model with nonlinear degradation, degradation rates $R_{26,i}$ and $R_{27,i}$ are replaced by

$$R_{26,i} = a_1[P^*S_i] + \frac{a_2[P^*S_i]}{[P^*S_i]^2 + a_3^2}, R_{27,i} = a_1[P^*S^*{}_i] + \frac{a_2[P^*S^*{}_i]}{[P^*S^*{}_i]^2 + a_3^2}$$

The empirical nonlinear second term is formulated in such a way that it approaches 0 at high concentrations of P*S and P*S*. This has an overall effect that the degradation of P*S and P*S* is faster at their lower concentrations to model the protective effect of STIL at high protein densities.

Values of all reaction rate constants $k_i$ are given in the Table S1. The system of ordinary differential equations that describes temporary dynamics of all model species in accordance with the reaction network shown in Figure 1 is as follows:

$$\frac{d[PS_i]}{dt} = -R_{1,i} + R_{-1,i} - R_{2,i} + R_{-2,i} - R_{7,i} - R_{8,i} + R_{10,i}\left(\frac{V_{cyt}}{V}, 1\right) + R_{15,i}\left(\frac{V_{cyt}}{V}, 1\right),$$
$$\frac{d[PS^*{}_i]}{dt} = R_{1,i} - R_{-1,i} - R_{3,i} + R_{-3,i} - R_{5,i} + R_{7,i} + R_{11,i}\left(\frac{V_{cyt}}{V}, 1\right),$$
$$\frac{d[P^*S_i]}{dt} = R_{2,i} - R_{-2,i} - R_{4,i} + R_{-4,i} - R_{6,i} + R_{8,i} - R_{26,i} + R_{12,i}\left(\frac{V_{cyt}}{V}, 1\right),$$
$$\frac{d[P^*S^*{}_i]}{dt} = R_{3,i} - R_{-3,i} + R_{4,i} - R_{-4,i} + R_{5,i} + R_{6,i} - R_{27,i} + R_{14,i}\left(\frac{V_{cyt}}{V}, 1\right),$$
$$\frac{d[P_i]}{dt} = -R_{15,i}\left(\frac{V_{cyt}}{V}, 1\right) + R_{16,i}\left(\frac{V_{cyt}}{V}, 1\right),$$
$$\frac{d[PS_c]}{dt} = -R_{1,c} + R_{-1,c} - R_{2,c} + R_{-2,c} - \sum_{i=1}^{N} R_{10,i}\left(1, \frac{V}{V_{cyt}}\right) + R_{15,c},$$
$$\frac{d[PS^*{}_c]}{dt} = R_{1,c} - R_{-1,c} - R_{3,c} + R_{-3,c} - \sum_{i=1}^{N} R_{11,i}\left(1, \frac{V}{V_{cyt}}\right),$$
$$\frac{d[P^*S_c]}{dt} = R_{2,c} - R_{-2,c} - R_{4,c} + R_{-4,c} - R_{26,c} - \sum_{i=1}^{N} R_{12,i}\left(1, \frac{V}{V_{cyt}}\right),$$
$$\frac{d[P^*S^*{}_c]}{dt} = R_{3,c} - R_{-3,c} + R_{4,c} - R_{-4,c} - R_{27,c} - \sum_{i=1}^{N} R_{14,i}\left(1, \frac{V}{V_{cyt}}\right),$$

$$\frac{d[P_c]}{dt} = R_{24} - R_{15,c} - \sum_{i=1}^{N} R_{16,i}\left(1, \frac{V}{V_{cyt}}\right),$$

$$\frac{d[S_c]}{dt} = R_{25} - R_{15,c} - \sum_{i=1}^{N} R_{15,i}\left(1, \frac{V}{V_{cyt}}\right),$$

here $V_{cyt}/V = 10^{-4}$ is the conversion factor (Milo and Phillips, 2016) due to the transition of molecules between the model compartments and should be substituted into the reaction rates determined in 4) and 5) instead of *a* and *b* in accordance with the above equations. To reduce the model complexity, we assume that only species P*S* can catalyse crossphosphorylation of PLK4 and STIL. However, no qualitative changes in the model behaviour were observed if other PLK4-STIL complexes were also permitted to catalyse the crossphosphorylation reactions. The molecular noise in our model is treated in the approach of chemical Langevin equation (Gillespie, 2000). Unless stated otherwise, noise amplitude in reaction $R_{15,i}$ was set to $am = 10^{-1}$. Application of noise to reactions other than reactions describing binding of proteins from the cytoplasm produces no qualitative differences in the model behavior. Initial concentrations of all species were equal to 0. PLK4 production was initiated at $t = 0$, while STIL production was initiated with the delay $t_{STIL} = 2h$. The system was solved numerically with the explicit super-time-stepping method (Alexiades et al., 1996) using a custom C code. Results of simulations were analyzed and plotted using Matlab (MathWorks, Natick, MA).

**Stability analysis of stationary states**

Stability analysis of the model symmetric stationary state (such as shown in Figure 3) was performed by numerically solving an eigenvalue problem of the order $5N + 6$. Due to this high order of the characteristic polynomial, the analysis was performed for the system with two components $N = 2$. To compute stability of the steady state on a 2D plane of parameters, as presented in Figure 3, the stationary state for the given values of chosen parameters was first found by the standard Newton method, using the previously computed values as the initial guess. Then, eigenvalues $\lambda_i$ were computed using the Arnoldi iteration method (Arnoldi, 1951). The region where max(Re($\lambda_i$)) is positive is shown in Figure 3 by color.

**Loss of symmetry breaking in the model without competition**

The system without communication between the centriolar compartments and, thus, competition, is defined by the following equation

$$k_{-10} = k_{-11} = k_{-12} = k_{-14} = k_{-15} = k_{-16} = 0, \quad [1]$$

i.e. reaction rate constants of dissociation of all model species from the centriole into the cytoplasm are set to zero as described in the main text. Substituting [1] into the model equations at the steady state one can readily obtain that the stationary concentration of the free PLK4 in all centriolar compartments are equal to

$$[P_i]_{SS} = \frac{k_{16}[P_c]_{SS}}{k_{15}[S_c]_{SS}}, \quad [2]$$

which implies that, regardless of the number of compartments, they all have equal concentration of free PLK4 that is determined only by the cytoplasmic concentrations of PLK4 and STIL. This is in contrast with the full model in which recycling of molecules back to the cytoplasm is enabled:

$$[P_i]_{SS} = \frac{k_{16}[P_c]_{SS} + k_{-15}V/V_{cyt}[PS_i]_{SS}}{k_{15}[S_c]_{SS} + k_{-16}V/V_{cyt}}. \quad [3]$$

As seen from equation [3], in this case, the concentration of free PLK4 is dependent on the concentration of PLK4-STIL in the same compartment and, thus, can be distinct for each model compartment. Furthermore, at the steady state the total degradation flux of STIL in each compartment,

$$k_{26}[P^*S_i]_{SS} + k_{27}[P^*S^*{}_i]_{SS}, \quad [4]$$

must be equal to the total influx of STIL from the cytoplasm

$$k_{10}[PS_c]_{SS} + k_{11}[PS^*{}_c]_{SS} + k_{12}[P^*S_c]_{SS} + k_{14}[P^*S^*{}_c]_{SS} \quad [5]$$

Since [5] is dependent only the global cytoplasmic concentration, this implies that all compartments have the same flux of STIL degradation and, therefore, PLK4. However, because [4] has two terms, there is a possibility to have an asymmetric state in which not all compartments have identical concentrations if and only if the compartments are locally bistable. Indeed, taking into the consideration [1], at a steady state the concentration of $[P^*S^*{}_i]_{SS} = X_i$ satisfies the following equation:

$$b_1X_i^4 + b_2X_i^3 + b_3X_i^2 + b_4X_i + b_5 = 0, \quad [6]$$

where all $b_i$ are functions only of the model parameters. Importantly,

$$b_1 < 0 \; and \; \; b_5 > 0$$

regardless of the values of model parameters. This implies that equation [6] can only have either 1 or 3 positive roots. In the first case the model has only one globally stable symmetric state (Figure 5A) and, thus, cannot exhibit symmetry breaking. In the second case, each compartment is bistable as in the model with nonlinear degradation considered in the main text. In this scenario, the system has three symmetric steady states (Figure 5C) of which only the lower symmetric state is accessible from the initial unpopulated state of the centriole as PLK4 and STIL are gradually produced in the cytoplasm.

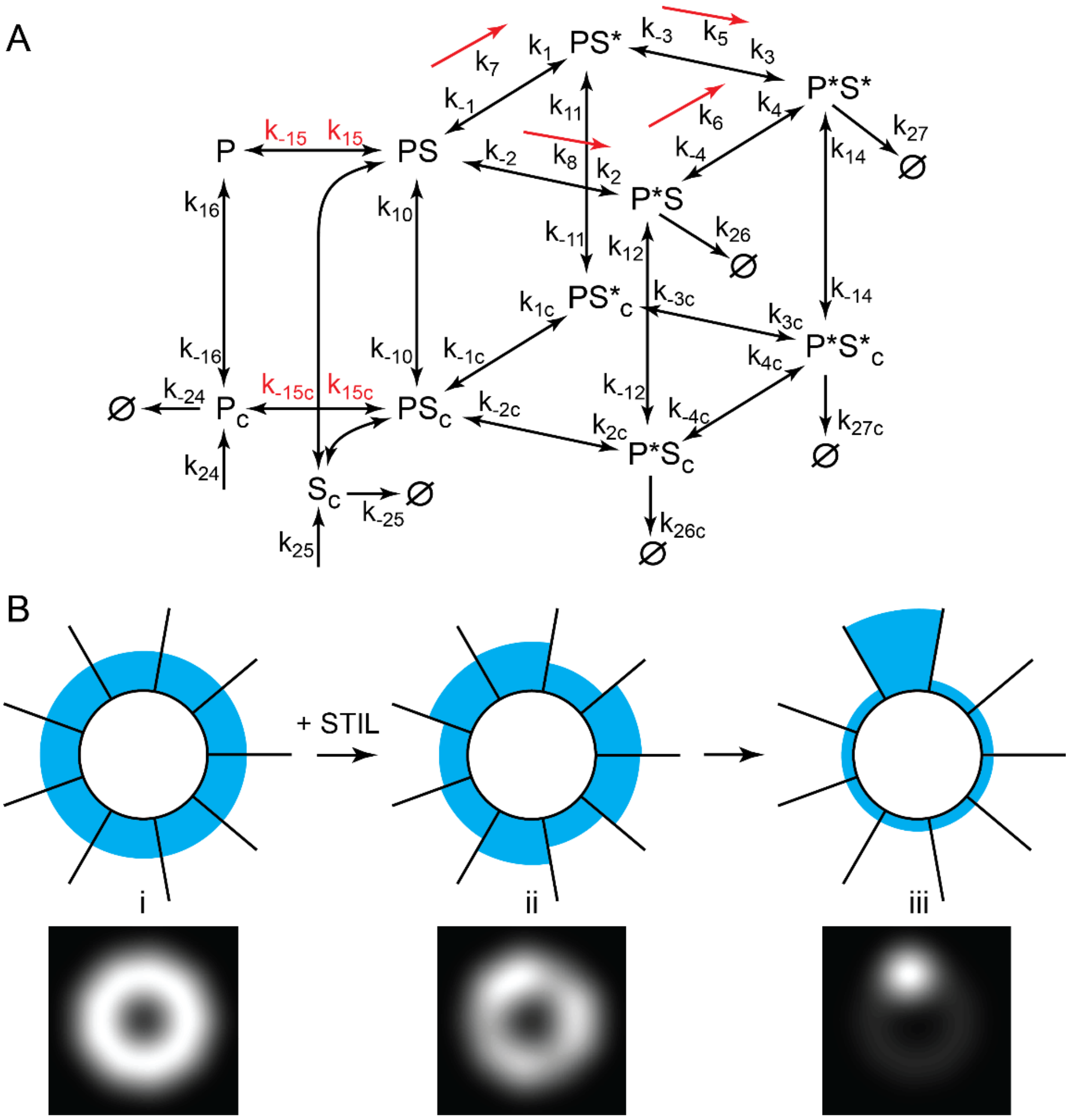

**Figure S1, related to Figures 1 and 2. Mathematical formulation of the model**. (A) Complete reaction diagram, including the reactions omitted for clarity in Figure 1. Reaction rate constants are shown next to the respective diagram arrows. Association and dissociation rates of complex formation reactions are shown only once (red font). Cross-phosphorylation reactions catalyzed by S*P*, e.g., SP + S*P* → S*P +S*P*, are shown schematically by red arrows. Numeric values of all reaction rates are given in Table S1. All other notations are the same as in Figure 1. (B) Spatial organization of the model. Upper row: model compartments are shown schematically as nine adjacent containers with PLK4 content shown by colored levels (not to scale). Bottom row: respective simulated fluorescence stills. Left to right: i) symmetric “ring” distribution of PLK4 around the centriole prior to the symmetry breaking; ii) just after the onset of symmetry breaking, on average half of compartments attempt to increase their PLK4 content; iii) a characteristic “spot” pattern of the PLK4 distribution upon the resolution of competition between the compartments. In the model, the compartments can exchange their content only via the common cytoplasm. Within the compartments all species are assumed to be spatially uniform (well-mixed).

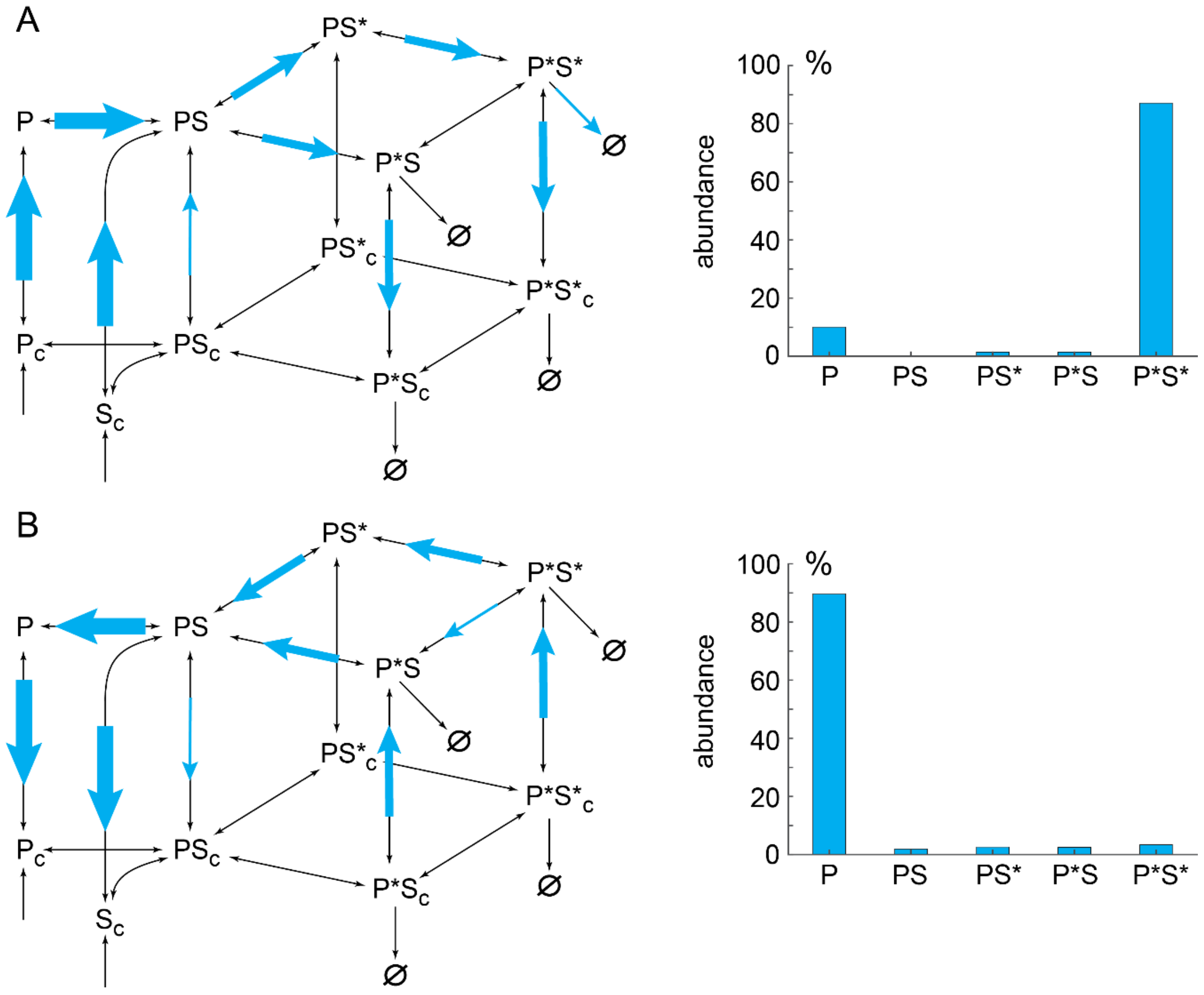

**Figure S2, related to Figures 2 and 4. Main reaction fluxes and relative abundances of centriole-bound protein complexes in the winning (A) and losing (B) compartments of the model ~2 hours after the symmetry breaking (5h after the start of simulation).** Left column. Relative magnitude of reaction fluxes is schematically shown by the size of colored arrows overlaid over the reaction diagram from Figure 1. The values of fluxes are given in Table S2. Right column. Relative abundances of PLK4 and its complexes with STIL are presented as bar plots.

## Supplemental tables

### Table S1. Model parameter values.

| Rate constant | Value | Reference |
|---|---|---|
| $k_1 = k_2 = k_3 = k_4$ | 1 $s^{-1}$ | this study |
| $k_{-1} = k_{-2} = k_{-3} = k_{-4}$ | 1 $s^{-1}$ | this study |
| $k_5 = k_6 = k_7 = k_8$ | 10 $(\mu M*s)^{-1}$ | this study |
| $k_{10} = k_{12} = k_{16}$ | 10 $s^{-1}$ | (Shimanovskaya et al., 2014) |
| $k_{-10} = k_{-12} = k_{-16}$ | 0.1 $s^{-1}$ | this study |
| $k_{11} = k_{14}$ | 10 $s^{-1}$ | this study |
| $k_{-11} = k_{-14}$ | 0.001 $s^{-1}$ | this study |
| $k_{15}$ | 10 $(\mu M*s)^{-1}$ | (Arquint et al., 2015) |
| $k_{-15}$ | 1 $s^{-1}$ | this study |
| $k_{26} = k_{27}$ | $10^{-4}$ $s^{-1}$ | (Klebba et al., 2015a; Klebba et al., 2015b) |
| $k_{1,c} = k_{2,c} = k_{3,c} = k_{4,c}$ | 1 $s^{-1}$ | this study |
| $k_{-1,c} = k_{-2,c} = k_{-3,c} = k_{-4,c}$ | 1 $s^{-1}$ | this study |
| $k_{15,c}$ | 10 $(\mu M*s)^{-1}$ | this study |
| $k_{-15,c}$ | 1 $s^{-1}$ | this study |
| $k_{24} = k_{25}$ | $10^{-7}$ $\mu M/s$ | this study |
| $k_{26,c} = k_{27,c}$ | $10^{-4}$ $s^{-1}$ | this study |
| $k_{-24} = k_{-25}$ | $10^{-5}$ $s^{-1}$ | this study |
| $V/V_{cyt}$ | $10^{-4}$ | this study |
| $a_1$ | 2 * $10^{-5}$ $s^{-1}$ | this study |
| $a_2$ | 4 * $10^{-5}$ $(\mu M)^2/s$ | this study |
| $a_3$ | 0.1 $\mu M$ | this study |

**Table S2. Numerical values of reaction fluxes in the winning and loosing compartments of the model, data for Figure S2. All values are in $10^{-3}\mu M \cdot s^{-1}$ .**

| **Reaction flux** | **Winning compartment** | **Losing compartment** |
|---|---|---|
| R16 | 11.59 | -1.302 |
| R15 | 11.57 | -1.328 |
| R1 | 6.386 | -0.575 |
| R2 | 6.495 | -0.9029 |
| R3 | 6.392 | -0.5438 |
| R4 | -0.1841 | -0.1162 |
| R10 | 1.353 | -0.1875 |
| R11 | -0.04353 | 0.03198 |
| R12 | -6.718 | 0.7893 |
| R14 | -5.237 | 0.6766 |
| R26 | 0.009834 | 0.002327 |
| R27 | 0.5944 | 0.003047 |

**Supplemental references**